\documentclass[a4paper,review,11pt,authoryear]{elsarticle}
\usepackage[a4paper, margin=1in]{geometry}

\usepackage{comment}
\usepackage{float}
\usepackage{ulem}
\usepackage{amssymb}
\usepackage{amsmath}
\usepackage[T1]{fontenc}
\usepackage{tikz}
\usepackage{mathtools}
\usepackage{natbib}
\usepackage{multirow}
\usepackage{makecell}
\usepackage{color}
\usepackage{soul}
\usepackage{graphicx}
\usepackage{caption}
\usepackage{xcolor}
\usepackage{algorithm}
\usepackage{algorithmicx}
\usepackage{algpseudocode}
\usepackage{enumitem}
\usepackage{lipsum}
\usepackage{caption} 
\usepackage[unicode=true]{hyperref}
\let\pkg=\texttt
\let\code=\texttt
\let\proglang=\textsf

\usepackage{booktabs}

\usepackage{longtable}
\usepackage[edges]{forest}
\usepackage{hyperref}
\hypersetup{
  colorlinks=true,
  linkcolor=blue,
  citecolor=blue,
  urlcolor=blue}
\usepackage{lineno}

\journal{}
\begin{document}
\begin{frontmatter}

\title{Combining Probabilistic Forecasts of Intermittent Demand}

\author[label1]{Shengjie Wang}
\ead{wsj19992017@buaa.edu.cn}
\author[label1]{Yanfei Kang\corref{cor1}}
\ead{yanfeikang@buaa.edu.cn}
\author[label2]{Fotios Petropoulos}
\ead{fotios@bath.edu}
\cortext[cor1]{Corresponding author}
\address[label1]{School of Economics and Management, Beihang University, China}
\address[label2]{School of Management, University of Bath, Bath, UK}

\begin{abstract}
In recent decades, new methods and approaches have been developed for forecasting intermittent demand series. However, the majority of research has focused on point forecasting, with little exploration into probabilistic intermittent demand forecasting. This is despite the fact that probabilistic forecasting is crucial for effective decision-making under uncertainty and inventory management. Additionally, most literature on this topic has focused solely on forecasting performance and has overlooked the inventory implications, which are directly relevant to intermittent demand. To address these gaps, this study aims to construct probabilistic forecasting combinations for intermittent demand while considering both forecasting accuracy and inventory control utility in obtaining combinations and evaluating forecasts. Our empirical findings demonstrate that combinations perform better than individual approaches for forecasting intermittent demand, but there is a trade-off between forecasting and inventory performance.
\end{abstract}


\end{frontmatter}

\newpage
\section{Introduction} \label{intro}

Intermittent demand refers to count data that includes a significant number of zero observations. Such demand patterns are prevalent in the field of inventory management, spanning various industries such as automotive \citep{jiang2021intermittent}, retailing \citep{sillanpaa2018forecasting}, and aerospace \citep{wang2016select}. According to \citet{nikolopoulos2021we}, intermittent demand is frequent among spare parts, comprising approximately 50\% of inventories. Additionally, items displaying intermittent demands play a critical role in the after-sale industry \citep{petropoulos2022forecasting}. They are commonly observed in granular hierarchical levels, including the daily demand of stock-keeping units (SKUs) \citep{fildes2019retail}. Therefore, accurately forecasting intermittent demand items is essential for improving inventory-related decision-making.

The main challenge in forecasting intermittent demand patterns is the presence of uncertainty in both the demand size and the interval between non-zero demand. This dual uncertainty makes it difficult to model intermittent demand data \citep{theodorou2021exploring}, although it accurately reflects complex economic processes \citep{bartezzaghi1999simulation}. Previous studies have mainly focused on point forecasts for intermittent demand, proposing various approaches such as parametric methods \citep[see, for example,][]{croston1972forecasting,syntetos2005accuracy,teunter2011intermittent}, non-parametric methods \citep[see, for example,][]{willemain2004new,nikolopoulos2011aggregate,petropoulos2015forecast}, and machine learning-based approaches \citep[see, for example,][]{jiang2021intermittent,lolli2017single,makridakis2021m5}. However, few studies have concentrated on probabilistic intermittent demand forecasting \citep{willemain2006forecast,snyder2012forecasting} and its evaluation \citep{kolassa2016evaluating}.

Research in the area of probabilistic intermittent demand forecasting remains relatively unexplored, at least in comparison to the extensive research on point forecasting. Similarly, there is a limited amount of research on intermittent demand forecast combinations, despite the widespread use of forecast combinations in economics \citep{hall2007combining,jore2010combining,del2016dynamic,mcalinn2019dynamic} and finance \citep{geweke2011optimal,kapetanios2015generalised}. Furthermore, most forecasting research focuses on forecasting accuracy or other forecasting metrics, while ignoring inventory implications, although intermittent demand forecasts are directly relevant to inventory management. In light of this, we agree with \citet{goltsos2021inventory} that a research gap exists when measuring forecasting performance and the inventory implications of the forecasts, and our study aims to partially fill this gap.

In this study, we construct probabilistic forecasting combinations for intermittent demand focusing on inventory management. The main contributions are: (1) applying probabilistic forecasting combinations, overlooked in previous research of intermittent demand, to improve forecasting, (2) explicitly considering inventory metrics both in constructing combinations and evaluating forecasts to increase the economic value of forecasting, and (3) improving the output of the forecasting process towards better inventory management, by using probabilistic forecasting combinations to achieve better probabilistic forecasts and considering inventory metrics to ameliorate inventory performance.

The rest of the paper is organized as follows. Section \ref{review} reviews the associated research on intermittent demand forecasting approaches and probabilistic forecasting combinations. Section \ref{individual} presents the different approaches to construct combinations of probabilistic forecasts, including approaches based on forecasting scoring rules and inventory cost. Section \ref{comb} introduces individual methods to produce probabilistic forecasts for intermittent demand. Forecasts from these methods will form a pool and will later be aggregated to form probabilistic forecast combinations. In Section \ref{empirical}, we use a real dataset from a recent forecasting competition to evaluate our proposed combinations. Section \ref{conclusion} concludes the paper.

\section{Related studies}\label{review}
\subsection{Forecasting for intermittent demand}

In this section, we review the literature on point (Section~\ref{sec:point}) and  probabilistic forecasting (Section~\ref{sec:prob}) of intermittent demand.

\subsubsection{Point forecasting for intermittent demand}\label{sec:point}
Parametric approaches to forecasting intermittent demand forecasting are based on separately forecasting the non-zero demand sizes and the inter-demand intervals. Croston's method \citep{croston1972forecasting} and its bias correction modifications proposed by \citet{syntetos2005accuracy}, also commonly known as Syntetos-Boylan Approximation (SBA), and \citet{shale2006forecasting} are popular parametric methods. To deal with inventory obsolescence, \citet{teunter2011intermittent} proposed a new method (Teunter-Syntetos-Babai or TSB) which focuses on updating the demand probability rather than the inter-demand interval. More recently, \citet{babai2019new} and \citet{yang2021modified} proposed further modifications to the TSB and SBA methods. 

\citet{syntetos2005categorization}, and later on \citet{kostenko2006note}, proposed a categorization of intermittent demand patterns based on the average demand interval (ADI) and the squared coefficient of variation of demand sizes (CV2). This classification has been a useful tool for analyzing intermittent demand items but has also been used to select between Croston's method and its SBA bias correction. It is suggested that smoother series (with lower CV2 and ADI) should be modeled by the original Croston's method. In contrast, the SBA method is more appropriate for more intermittent, erratic and lumpy series.

Apart from the aforementioned parametric methods, non-parametric methods that do not make explicit assumptions regarding the distributions of the data have also been proposed in the literature. Two very popular categories of non-parametric methods are bootstrapping and temporal aggregation. Bootstrapping methods are based on resampling the historical data to construct empirical demand distributions \citet{willemain2004new,zhou2011comparison}. Temporal aggregation methods focus on transforming the original series to new series of lower frequency, and thus of lower intermittence. \citet{nikolopoulos2011aggregate} first proposed the Aggregate-Disaggregate Intermittent Demand Approach (ADIDA). Later, \citet{petropoulos2015forecast} considered combinations of forecasts with different frequencies and called their approach the Intermittent Multiple Aggregation Prediction Algorithm (IMAPA). To capture the hidden trend and seasonal patterns in the data, \citet{kourentzes2021elucidate} introduced temporal hierarchies for intermittent demand series.

Some studies have applied machine learning methods for intermittent demand forecasting, such as support vector machines \citep{jiang2021intermittent} and  neural networks \citep{lolli2017single}. Tree-based machine learning methods like the light gradient-boosting machine (LightGBM) performed exceptionally in the M5 forecasting competition \citep{makridakis2021m5,makridakis2022m5}, especially for the higher (less granular) hierarchical levels.

\subsubsection{Probabilistic forecasting for intermittent demand}\label{sec:prob}

The methods mentioned above focus on producing point forecasts and primarily target forecasting performance (accuracy and bias). Previous studies have shown that these methods can improve intermittent demand forecasting while being diverse in nature and thus potentially useful for forecast combinations. On the other hand, point forecasts focusing on mean or median may not be sufficient to cope with many decision-making problems, such as inventory management, since point forecasts do not provide information related to the uncertainty around the forecasts. \citet{kolassa2016evaluating} stressed that producing forecasts for the whole distribution is meaningful, particularly for supply chain forecasting and inventory control applications.

Several studies have focused on probabilistic intermittent demand and how to measure their performance. \citet{willemain2006forecast} suggested forecasting the entire demand distribution and evaluating the accuracy by using a $\chi^2$ statistic, assuming that distributions are independent and identical over time. \citet{snyder2012forecasting} applied different models to fit distributions of intermittent demand and proposed prediction likelihood score (PLS) and used the discrete ranked probability score (DRPS) to evaluate their predictive distributions. \citet{kolassa2016evaluating} further discussed the evaluation of probabilistic forecasting in the context of intermittent demand. \citet{sillanpaa2018forecasting} found that using probabilistic forecasts directly in inventory policy could provide better performance than point forecasts. \citet{trapero2019quantile} combined GARCH models and kernel density estimation to estimate safety stocks. Finally, there has been some controversy on whether distributions should be produced for intermittent demand forecasting. \citet{boylan2006accuracy} noted that evaluating the accuracy of the whole distribution can result in misleading outcomes and that for inventory-related applications, quantiles towards the upper end of the distribution may be more appropriate. \citet{kolassa2016evaluating} argued that the entire distribution should be considered as a single quantile may not be enough.

\subsection{Probabilistic forecasting combinations}

This section focuses on the general probabilistic forecasting combinations. Combining point or probabilistic forecasts produced by a variety of methods has been shown to outperform individual forecasts, with many studies, especially in the economic and finance domains, demonstrating significant performance benefits \citep[][\S2.6.1 and \S2.6.2]{petropoulos2022forecasting}. A simple and useful combination is the linear combination of finite individual forecasts with non-negative weights that sum to a unit. The key challenge is how to distribute or estimate weights for each individual forecast method.

One option would be to assign equal weights for each individual method. Equal weights (arithmetic mean) are a straightforward solution that usually results in a robust outcome that can also act as a benchmark for comparing the performance of other non-equally weighted combinations. \citet{wallis2005combining} and \citet{o2006uncertain} offer reviews of the performance of the equal-weight or simple average probabilistic combinations.

Towards achieving ``optimal'' weights to improve linear combinations, several researchers proposed various optimization approaches. \citet{hall2007combining} suggested estimating weights by minimizing the Kullback-Leibler information criterion (KLIC), which is the distance between the combined distribution and the real but unknown distribution. They gave an equivalent form of maximizing the logarithmic score of the combined distribution. \citet{jore2010combining} used sliding windows of historical data and constructed weights in terms of logarithmic scores based on the past performance of each individual method. \citet{geweke2011optimal} offered a theoretical analysis to support probabilistic combinations based on the logarithmic score. \citet{conflitti2015optimal} supplied an iteration algorithm to simplify optimization. \citet{kapetanios2015generalised} proposed the ``generalized pool'' to allow time-varying optimal weights. \citep{del2016dynamic} constructed ``dynamic prediction pools'' to simulate the varying process of weights. \citet{li2022bayesian} estimated time-varying weights based on time-varying features. Apart from the logarithmic score, the censored likelihood scoring rule \citep{diks2011likelihood,opschoor2017combining} and the continuously ranked probability score \citep{gneiting2014probabilistic} can also be used towards estimating weights for probabilistic forecast combinations.

Except for linear combinations, probabilistic combinations include other techniques such as nonlinear pools \citep{gneiting2013combining,ranjan2010combining,bassetti2018bayesian}, Bayesian model averaging (BMA) \citep{garratt2003forecast,moral2015model,aastveit2018evolution}, Bayesian predictive synthesis (BPS) \citep{mcalinn2019dynamic,mcalinn2020multivariate} and quantile forecasting \citep{lichtendahl2013better,busetti2017quantile,trapero2019quantile}. 
These probabilistic combination approaches have specific advantages and shortcomings. The linear pool is simple enough to understand and optimize but may become under-confident because some combinations may lead to larger variances than the linear combination of individual variances \citep{ranjan2010combining}. The nonlinear pool attempts to solve this issue. However, there is no evidence of significant differences between the performance of linear and nonlinear pools \citep{baran2018combining}. BMA offers an elegant theoretical approach to model uncertainty \citep{garratt2003forecast} but is based on the assumption that the true model is amongst the considered pool of model \citep{wright2008bayesian}. BPS aims to deal with the correlation among different individual forecasts. However, the approach is overall more complex than the previous ones. For more details about combination approaches, the reader is referred to the overview article by \citet{wang2022forecast}. 

Notably, among these studies, the utilized data are primarily restricted to the economic and financial contexts, depending on the academic background and interest of the researchers. We believe it would be valuable to apply these combination approaches to other fields, such as intermittent demand, where research on probabilistic combinations is rare. In addition, existing studies on forecasting combinations focus more on the accuracy of forecasts and less on the practical implications, such as the inventory performance of intermittent demand items. In our work, we construct probabilistic forecast combinations, considering both the forecasting accuracy and the inventory performance. We compare the outcomes of different combinations to provide insights for improving intermittent demand forecasting and inventory management.

\section{Individual probabilistic forecasting of intermittent demand} \label{individual}

In this section, we consider five mainstream types of individual forecasting models/methods in the intermittent demand forecasting literature. They are based on quantile regression, distributions, bootstrap, machine learning and traditional time series models, respectively. In total, we consider nine individual methods to be combined. The two machine-learning-based methods are global~\citep{bandara2021improving}, meaning that all the series learn across each other in the training process, while the other seven individual models/methods are local. Table \ref{tab:intro-individual} briefly describes them.

\subsection{Quantile-regression-based models}\label{qr}

Quantile regression is a widely employed and valuable approach to achieving quantile forecasts and particularly adept at incorporating exogenous information, such as holidays, seasonality, and policies. For example, \citet{gaillard2016additive}  successfully utilized quantile regression in probabilistic electricity load and price forecasting and won the first prize in Global Energy Forecasting Competition 2014~\citep{HONG2016896}.

Motivated by \citet{gaillard2016additive}, in the context of intermittent demand forecasting, we employ count quantile regression \citep{machado2005quantiles} with generalized additive models (GAMs) \citep{hastie1990generalized}. The pioneering use of quantile regression in forecasting was initiated by 
\citet{taylor2000quantile}, who introduced a new method for estimating conditional density. The generalized additive model was introduced by  \citet{hastie1990generalized}, which can capture the non-linear relationships between the demand mean and the covariates in our forecasting.

\begin{enumerate}
    \item \textbf{Fitting GAM}. The GAM model aims to model the conditional mean of $Y_t$ as
\begin{equation}
    g(E(Y_t|X_{t, 1}, \cdots, X_{t,P}))=f_1(X_{t,1})+\cdots+f_M(X_{t,P}),
    \label{eq:gam}
\end{equation}
where $Y_t$ denotes the demand at time $t$, $X_{t,1},\cdots,X_{t,P}$ denotes the $P$ covariates at time $t$. $g(\cdot)$ is a link function where the logarithmic function is applied. $f_1(\cdot),\cdots,f_P(\cdot)$ are the transformation functions of the covariates, with the form of cubic splines for continuous ones and the identity function for others such as categorical variables and retail prices with little variation.  We use \proglang{R} Packages \pkg{mgcv} \citep{Wood2016} to fit the GAM model.

    \item \textbf{Fitting count quantile regression}. We perform count quantile regression with the estimated components $\hat{f}_1(\cdot),\cdots,\hat{f}_P(\cdot)$ for quantile levels $\tau \in \{0.01,0.02,\cdots,0.99\}$ to forecast quantiles:
    \begin{equation}
    Q_{T(Z,\tau)}(\tau|\boldsymbol{\hat{f}})=\boldsymbol{\hat{f^{'}}} \boldsymbol{\beta(\tau)},
    \label{eq:count-quan}
    \end{equation}
    where  $\boldsymbol{\hat{f}}=(\hat{f}_1(\cdot),\cdots,\hat{f}_P(\cdot))$, and $Z=Y+U$, where $U$ is a uniform in $[0,1)$ random variable,  independent of $\boldsymbol{\hat{f}}$ and $Y$. $T(Z,\tau)$ is a transformation function:
      \begin{equation*}
    \label{eq:trans}
    T(Z,\tau)=\left\{
    \begin{aligned}
    {\rm log}(Z-\tau) & , & Z>\tau,\\
    {\rm log}(10^{-5})  & , & Z \leq \tau.
    \end{aligned}
    \right.
    \end{equation*}
    Equation \eqref{eq:count-quan} performs a quantile regression for $T(Z,\tau)$ at quantile $\tau$ based on  $\boldsymbol{\hat{f}}$. 
     The estimate of the target quantile $Q_{Y}(\tau|\boldsymbol{\hat{f}})$ is equal to $\lceil Q_{T(Z,\tau)}(\tau|\boldsymbol{\hat{f}})-1 \rceil$. We use the \proglang{R} package \pkg{quantreg} \citep{Koenker2022} for count quantile regression.
\end{enumerate}

\subsection{Distribution-based models}\label{distribution}

Distribution-based models aim to estimate specific distributions of intermittent demand with statistical models. While the Poisson distribution is commonly used for count data, its popular extension, the negative binomial distribution, can cope with overdispersion. Therefore, in this paper, we consider Poisson distribution and negative binomial distribution, with the distribution mean, $\mu_t$, varying following a damped dynamic relationship~\citep{snyder2012forecasting} to model the autocorrelation of the demand.
\begin{equation}
    \mu_t=(1-\phi-\alpha)\mu+\phi\mu_{t-1}+\alpha{y_{t-1}},
    \label{eq:damped}
\end{equation}
where the long-run mean $\mu$, $\phi$, and $\alpha$ are all positive, with $\phi+\alpha<1$. $y_t$ is the observation at time $t$. Details of parameter settings can refer to \citet{snyder2012forecasting}.

For the two distribution-based models, parameters are optimized with maximum likelihood estimation (MLE) by using \proglang{R} package \pkg{maxLik} \citep{henningsen2011maxlik}. We simulate the future series 1,000 times based on the estimated parameters and calculate the corresponding quantiles accordingly.

\subsection{Bootstrap-based method}\label{bootstrap}

The bootstrap method \citep{eforn1979bootstrap} constructs an empirical distribution by resampling the historical demand data. There has been an amount of related research on probabilistic intermittent demand forecasting, including \citet{willemain2004new}, \citet{zhou2011comparison}, and \citet{hasni2019performance}. This paper considers the two bootstrap-based methods from \citet{willemain2004new} and \citet{zhou2011comparison}, named WSS and ZV, respectively. WSS approach estimates the probability of zero/non-zero demand based on a two-state Markov chain. It samples from the estimated probability to define where the positive demands are in the lead time, and estimates the size of positive demand by sampling historical data with an extra random item. ZV approach samples lengths of demand intervals and sizes of positive demands to build the empirical distributions during lead time. Each method samples the future series 1,000 times in the forecasting process, and the corresponding quantiles are calculated accordingly.

\subsection{Machine-learning-based methods}

Machine learning methods are another popular stream in forecasting intermittent demand. For example, DeepAR \citep{salinas2020deepar} contributes a solution to probabilistic demand forecasting by training an autoregressive recurrent neural network model on a large collection of related time series. Besides, in both the accuracy \citep{makridakis2022m5} and uncertainty \citep{makridakis2021m5} tracks of M5 competition, machine learning methods such as LightGBM \citep{ke2017lightgbm} and LSTM (Long Short-Term Memory) \citep{hochreiter1997long,graves2012long}, achieved competitive performances. In this research, we use LightGBM and LSTM as two machine-learning-based individual methods to be combined for their outstanding performance in recent forecasting practice. LightGBM is an improved algorithm based on Gradient Boosting Decision Tree (GBDT) \citep{friedman2001greedy}, which reduces training complexity and is suitable for big data. LSTM can solve problems of vanishing gradients in the training process and model time series explicitly, which adapts well to time series forecasting. 

LightGBM estimates the quantiles concerning $\tau \in \{0.01,0.02,\cdots,0.99\}$, with the pinball loss as the loss function:
\begin{equation*}
    L_{\tau}(\hat{y},y)=\tau(\hat{y}-y)\boldsymbol{1}[\hat{y}>y]+(1-\tau)(y-\hat{y})\boldsymbol{1}[\hat{y}\leq y],
    \label{eq:pinball}
\end{equation*}
where $\hat{y}$ is the estimated quantiles and $y$ is the realizations. The $\boldsymbol{1}[\cdot]$ is the indicator   For LightGBM, forecasts are attained by the \proglang{Python} module \pkg{lightgbm}~\citep{ke2017lightgbm}. The specific hyperparameters settings can be seen in Table A.1 of the online Appendix A; other settings keep default values.

LSTM estimates the two parameters of negative binomial distributions with the loss function of negative likelihood. In LSTM, a simple network with five layers is designed. The first two layers consist of an embedding layer with a dropout layer to reduce the dimension of inputs. The reduced category feature dimensions can be found in Table A.2 of the online Appendix A. The following two are LSTM layer with a dropout layer, where the input size of the LSTM layer is 71, and the output size is 142, containing two hidden layers. The last one is a linear layer for the output of two distribution parameters. The drop probability of all dropout layers is 0.1. Other settings also preserve the default in the  \proglang{Python} module \pkg{Pytorch}
~\citep{paszke2019pytorch}.

\subsection{Traditional time series models}\label{tradition}

In addition to the aforementioned individual approaches, two traditional time series models, ARIMA \citep{forecast1} and ETS \citep{hyndman2002state}, are included in forecasting combinations. They were both used in the uncertainty track of M5 competition as benchmarks. We use the \proglang{R} package \pkg{forecast} \citep{forecast1} to generate probabilistic forecasts with the same settings as used in M5 competition. Point forecasts are attained first, and normal distributions are constructed based on point forecasts and forecasting errors. Given the non-negative nature of demand in practice, the negative parts of distributions are truncated as the probability of zero. 

\begin{table}[ht!]
\centering
\caption{A brief description of the nine individual models/methods to be combined.}
\label{tab:intro-individual}
\resizebox{\columnwidth}{!}{\begin{tabular}{lp{0.15\columnwidth}p{0.65\columnwidth}}
\toprule
Type & Name & Description \\ \midrule 
\multirow{2}*{Quantile regression} & \multirow{2}*{GAM-QR} & A quantile regression model with GAM to extract nonlinearity~\citep{gaillard2016additive}.\\\hline
\multirow{2}*{Distributions} & POIS & A Poisson model with its mean varying (see Equation (\ref{eq:damped})). \\
~ & NB & A negative binomial model with its mean  varying (see Equation (\ref{eq:damped})). \\\hline
\multirow{2}*{Bootstrap} & WSS & A Bootstrap method proposed by \citet{willemain2004new}. \\
~ & ZV & A Bootstrap method proposed by \citet{zhou2011comparison}. \\\hline
\multirow{2}*{Machine Learning} & LightGBM & LightGBM method used to estimate quantiles. \\
~ & LSTM & LSTM method used to estimate a negative binomial distribution.\\\hline
\multirow{2}*{Traditional} & ARIMA & ARIMA family of models. \\
~ & ETS & The exponential smoothing state space family of models. \\
\bottomrule 
\end{tabular}}
\end{table}

\section{Probabilistic forecast combination in intermittent demand}\label{comb}

Probabilistic forecasts can be presented in various forms, such as densities, distributions, quantiles, and prediction intervals. As for intermittent demand, according to \citet{kolassa2016evaluating}, distribution may adapt to more complicated decisions on inventory management because of containing more information than quantiles. In addition, combining quantiles may introduce non-integer support, which violates the properties of intermittent demand as a type of count data. However, certain methods, such as quantile regression, are designed to yield quantiles. So for the unity of our proposed individual methods, we first attain the quantile forecasts, and subsequently transform them into distributions as the final outputs.

For simplicity, we employ the linear pooling to combine individual probability forecasts. Suppose that there are $N$ individual forecasts with cumulative distribution functions (CDFs) of a random variable $Y$ at time $T+h$ and information set $I$ available until time $T$, denoted $F_i^{(T+h)}(y|I_T)$, $i=1,\cdots, N$. Then the combination forecasts can be obtained by calculating the finite mixture
\begin{equation}
    F_{\text{comb}}^{(T+h)}(y|I_T)=\sum^{N}_{i=1}w_{T+h|T,i}F_i^{(T+h)}(y|I_T),
    \label{eq:combination definition}
\end{equation}
where $w_{T+h|T,i}$ is the weight of the $i$th individual probability forecast $F_i^{(T+h)}(y|I_T)$. To ensure that the combination forecasts maintain the properties of probability distributions, the weights $w_{T+h|T,i}$ are constrained to be non-negative and sum to one. Besides, the weights remain constant over time in the paper for simplicity. 

The key to the linear pooling approach lies in optimizing the mixing weights. Consequently, three distinct methods are devised to obtain these weights, including distributed weights, scoring-rule-based optimal weights and inventory-cost-based optimal weights. In total, we explore seven combination approaches, each briefly described in Table \ref{tab:intro-combination}.

\subsection{Simple weights}

A simple way to assign the weights is to set all weights equal to $1/N$, also called simple averaging. It can often achieve robust and stable outcomes. Similar to point forecasts, simple averaging (SA) is commonly challenging to beat in probabilistic forecasts~\citep{wang2022forecast}. Therefore, SA is used as a crucial combination benchmark in our paper.

Another weighting scheme is based on historical performance, represented by logarithmic scores in this study.  We calculate the combining weights as follows:
\begin{equation}
    {\rm logscore}_i=\frac{1}{h}\sum^{T}_{t=T-h+1}\log(f_{\text{comb}}^{(t)}(y_{t}|I_{T-h})),
    \label{eq:log score1}
\end{equation}
\begin{equation}
    w_{T+h|T,i}=\frac{1}{{\rm logscore}_i}/\sum_{j=1}^N (\frac{1}{{\rm logscore}_j}),
    \label{eq:past log score weight}
\end{equation}
where ${\rm logscore}_i$ gives the average logarithmic scores for the $i$th individual forecast during the past $h$ periods, $f_{\text{comb}}^{(t+h)}(y|I_t)$ represents the combined probability mass function (PMF) forecasts, and $w_{T+h|T,i}$ calculates the combining weights for the $i$-th individual forecast. 

\subsection{Scoring-rule-based optimal weights}
In this section, we consider four scoring rules, namely logarithmic scores, censored likelihood (CL), Brier score, and discrete ranked probability score (DRPS) to optimize the combining weights. Each scoring rule is briefly described as follows.

\subsubsection*{Optimizing logarithmic scores}
  
  \citet{hall2007combining} proposed to obtain combining weights by optimizing the logarithmic score \citep{mitchell2005evaluating} as follows:
\begin{equation}
    \boldsymbol{w_{T+h|T}}=
{ \underset {\boldsymbol{w}} { \operatorname {arg\,max} } \, \frac{1}{h}\sum^{T}_{t=T-h+1}\log(f_{\text{comb}}^{(t)}(y_{t}|I_{T-h+1})) },
    \label{eq:log score weight}
\end{equation}
where the weights of forecasting combination $\boldsymbol{w_{T+h|T}}=(w_{T+h|T,1},\cdots,w_{T+h|T,N})^{\prime}$. This method is derived from the minimization problem of the Kullback-Leibler information criterion (KLIC) distance between the combined distribution and the true but unknown distribution. The initial approach we employ to obtain optimal weights involves optimizing the average historical logarithmic score.

\subsubsection*{Optimizing censored likelihood (CL) scoring rule}

 Concerning intermittent demand, focusing more on upper quantiles seems vital because they are more valuable in inventory management, such as planning safety stock. \citet{boylan2006accuracy} point out that concentrating on the whole distribution may yield misleading information, resulting in suboptimal performance for forecasts at higher quantiles. Considering that, we apply CL scoring rule \citep{diks2011likelihood}, a variant of the logarithmic score. The score has previously been utilized by \citet{opschoor2017combining} in the estimation of Value at Risk, also requiring paying more attention to extreme quantiles. The CL score proves to be valuable in situations where the focus is on a specific area of a probabilistic distribution. For a combined probabilistic forecast with the specific area $A$, its CL score can be denoted as:
\begin{align*}
      \text{CL}(f_{\text{comb}}^{(t+h)}(y|I_t))=& \boldsymbol{1}[y_{t+h}\in A_{t+h}] \log(f_{\text{comb}}^{(t+h)}(y_{t+h}|I_t)) \\
    &+ \boldsymbol{1}[y_{t+h}\in A_{t+h}^C] \log(\int_{A_{t+h}^C} f_{\text{comb}}^{(t+h)}(y|I_t) dy),
    \label{eq:csl}
\end{align*}
with $A_{t+h}^C$ the complement of $A_{t+h}$ and $\boldsymbol{1}(\cdot)$ the indicator function, which is equal to 1 when the argument is true and 0 otherwise. The score captures the shape of the distribution in area $A_{t+h}$. In this paper, the region $A_{t+h}$ is defined as values $y_{t+h}$ greater than the 90\%th quantile of the forecast distributions, emphasizing high sales of intermittent demand. Optimizing the CL score is our second approach to obtaining the optimal weights. That is to say, the weights are yield as Equation (\ref{eq:cl weight}), similar to optimizing logarithmic scores:

\begin{equation}
    \boldsymbol{w_{T+h|T}}=
{ \underset {\boldsymbol{w}} { \operatorname {arg\,max} } \, \frac{1}{h}\sum^{T}_{t=T-h+1}\text{CL}(f_{\text{comb}}^{(t)}(y_{t}|I_{T-h+1})) }.
    \label{eq:cl weight}
\end{equation}

We apply an iterative scheme to optimize the two scores above~\citep{conflitti2015optimal}. For the $i$-th individual method, the iteration of the corresponding weight based on optimizing the logarithmic score is performed as follows:
\begin{equation}
    w_{T+h|T,i}^{(k+1)}=w_{T+h|T,i}^{(k)}\frac{1}{h}\sum_{t=T-h+1}^{T}\frac{\hat{f}_{i}(y_{t}|I_{T-h})}{\sum_{j=1}^{N}\hat{f}_{j}(y_{t}|I_{T-h})w_{T+h|T,j}^{(k)}},
    \label{eq:opt-log}
\end{equation}
where $\hat{f}_{i}(y_{t}|I_{T-h})$ is the estimated probability of method $i$ in the forecasting period from $T-h+1$ to $T$. When optimizing the CL score, the item $\hat{f}_{i}(y_{t}|I_{T-h})$ is replaced by $\boldsymbol{1}[y_{t}\in A_{t}]\hat{f}_{i}(y_{t}|I_{T-h})+\boldsymbol{1}[y_{t}\in A_{t}^C]\int_{A_{t}^C} f_{i}(y|I_{T-h}) dy$.  The optimization starts with equal weights and uses a tolerance of 0.001 of the Euclidean distance of two successive iterations as a stopping criterion.

\subsubsection*{Optimizing Brier score}
Another scoring rule for optimization is the Brier score \citep{brier1950verification}. It was introduced by \citet{kolassa2016evaluating} as a forecasting evaluation metric for intermittent demand. It can be denoted as 
\begin{equation}
    {\rm Brier}(f_{\text{comb}}^{(t+h)}(y|I_t)) = -2f_{\text{comb}}^{(t+h)}(y_{t+h}|I_t)+\sum_{k=0}^{\infty} f_{\text{comb}}^{(t+h)}(k|I_t)^2,
    \label{eq:brier}
\end{equation}
where $f_{\text{comb}}^{(t+h)}(k|I_t)$ is the value of the probability mass function $f_{\text{comb}}^{(t+h)}(y|I_t)$ at the value $k$. We calculate weights by minimizing the Brier score as Equation (\ref{eq:brier weight}), an optimization problem of the quadratic target function.
\begin{align}    
    \boldsymbol{w_{T+h|T}}=
{ \underset {\boldsymbol{w}} { \operatorname {arg\,min} } \, \frac{1}{h}\sum^{T}_{t=T-h+1}\text{Brier}(f_{\text{comb}}^{(t)}(y_{t}|I_{T-h+1}))}.
    \label{eq:brier weight}
\end{align}

\subsubsection*{Optimizing discrete ranked probability score (DRPS)}
In addition, we consider DRPS as an optimization objective to yield weights. DRPS was utilized by \citet{snyder2012forecasting} to evaluate the forecast distribution of intermittent demand. It can be denoted as
\begin{equation}
    {\rm DRPS}(F_{\text{comb}}^{(t+h)}(y|I_t),y_{t+h})=\sum_{k=0}^{\infty}(F_{\text{comb}}^{(t+h)}(k|I_t)-\boldsymbol{1}[y_{t+h} \leq k])^2,
    \label{eq:drps}
\end{equation}
where $F_{\text{comb}}^{(t+h)}(k|I_t)$ is the value of cumulative probability $F_{\text{comb}}^{(t+h)}(y|I_t)$ at $k$. The weights can be achieved by minimizing DRPS:
\begin{equation}
    \boldsymbol{w_{T+h|T}}=
{ \underset {\boldsymbol{w}} { \operatorname {arg\,min} } \, \frac{1}{h}\sum^{T}_{t=T-h+1}\text{DRPS}(f_{\text{comb}}^{(t+h)}(y_{t}|I_{T-h+1})) }.
    \label{eq:drps weight}
\end{equation}
The solution can be achieved easily due to the quadratic target function, resembling the Brier score rule. We execute the optimization of Brier scores and DRPS scores using \proglang{R} function \code{constrOptim()}.

\subsection{Inventory-cost-based optimal weights}\label{comb-cost}

Before introducing the optimization objective, we provide an overview of the presumed inventory policy, which dictates how forecasts are employed in inventory management and then influences costs. We first set a target customer service level (CSL) $\tau$ as the designated probability of the forecasting distribution. Then the corresponding quantiles are calculated as inventory holdings. Eventually, the cost from inventory holdings charges and lost sales is evaluated regarding differences between forecasts and real demand. Given the inventory policy, optimal weights based on inventory cost functions can be yielded as follows.
\begin{equation}
    \boldsymbol{w_{T+h|T}}=
{ \underset {\boldsymbol{w}} { \operatorname {arg\,min} } \,\text{cost}=\underset {\boldsymbol{w}} { \operatorname {arg\,min} } 
 \left[c_1*\text{stock}+c_2*\text{lostsales}\right]},
    \label{eq:obj}
\end{equation}
where $\text{stock} =\frac{1}{h}\sum_{t=T-h+1}^{T} [\hat{q}_t(\tau)-x_t]_+$ and $\text{lostsales}=\frac{1}{h}\sum_{t=T-h+1}^{T} [x_t-\hat{q}_t(\tau)]_+$, and
$\tau=\frac{c_2}{c_2+c_1}$ shows different proportional relations, where $c_1$ and $c_2$ are the marginal costs of inventory holding charges and lost sales. $x_t$ is the real sales at time $t$. The estimated quantile $\hat{q}_t(\tau)$ at the specific target level $\tau$ of the combined distribution $F_{\text{comb}}$ meets $F_{\text{comb}}(\hat{q}_t(\tau))\geq \tau$ and $F_{\text{comb}}(\hat{q}_t(\tau)-1)< \tau$. The quantiles are calculated from the combined distributions as Equation (\ref{eq:combination definition}), linked to optimal weights. Given the two parameters of marginal cost, the target customer service level can be defined as $\tau=\frac{c_2}{c_2+c_1}$ according to the newsvendor problem \citep{huber2019data}. To simulate varying proportions of marginal costs between stock and lost sales, corresponding to the target level of 80\%, 90\% and 95\%, we explore three groups of choices for $c_1$ and $c_2$. They are (1, 4), (1, 9) and (1,19), where (1, 4) means $c_1 = 1$, and $c_2 = 4$. 

Note that weights are contained in the combined distribution, and the distribution is discrete. Therefore, the cost function is discontinuous. Moreover, the optimal problem may be non-convex due to the complexity of the inventory function, as highlighted by \citet{kourentzes2020optimising}. Given these challenges, Particle Swarm Optimization (PSO) \citep{kennedy1995particle} is applied to solve the optimization problem.

\begin{table}[!ht]
\centering
\caption{A brief description of the combination methods.}
\label{tab:intro-combination}
\resizebox{\columnwidth}{!}{\begin{tabular}{llp{0.7\textwidth}}
\toprule
Type & Name & Description\\ \midrule 
\multirow{2}*{Simple weights} & SA &The simple average of all individual methods.\\
~ & log-score & Distributed weights based on historic logarithmic scores.\\\hline
\multirow{4}*{Scoring-rule-based optimal weights}& log-opt & Optimal weights based on maximizing logarithmic scores. \\
~ & cl-opt &Optimal weights based on maximizing CL scores.\\
~ & brier-opt & Optimal weights based on minimizing Brier scores.\\
~ & drps-opt & Optimal weights based on minimizing DRPS scores. \\\hline
\multirow{2}*{Inventory-cost-based optimal weights} & \multirow{2}*{cost$c_2$-opt} & Optimal weights based on minimizing inventory costs  with (1, 4), (1, 9) and (1,19) for $c_1$ and $c_2$. \\

\bottomrule 
\end{tabular}}
\end{table}

\section{Empirical evaluation}\label{empirical}

\subsection{Data description and preparation}

We apply the proposed method to M5 competition dataset~\citep{makridakis2022m5}. M5 data comprises 3049 units of products sold by Walmart at ten stores in the USA from January 29, 2011 to May 23, 2016 (1941 days). The dataset exhibits a complex hierarchical structure. We only consider the bottom level in this paper. Approximately 60.1\% of sales are zeros, with counting commencing from the first positive value for every series, presenting obvious intermittence.

Aligned with the design of M5 competition, we examine 28-day-ahead forecasts in this paper. To facilitate the training and evaluation of forecasts, we split each series into three parts. The first one spans from the first period with positive sales to the last 57th period, utilized for training individual methods and forecasting the next 28 periods. Exceptions are LightGBM and LSTM, which only use data in the recent two years due to their computational costs. The second part encompasses the period from the 56th to the 29th for estimating the combination weights. The third part consists of the last 28 periods of the whole series for evaluating the proposed combination approaches. The bottom level of M5 data includes the sales time series of 30,490 SKUs (Stock Keeping Units). 1587 SKUs exhibit consistent zero demand in the second part, i.e., the weight estimation period. We exclude these SKUs in this study since the all-zero demand introduces a potential distortion in combinations. For instance, forecasting methods leaning towards zero may be disproportionately allocated larger weights, resulting in significant errors if confronted with considerable positive values during evaluation periods. It may not be appropriate to evaluate combinations using extreme data.  Subsequently, we consider 28,903 SKUs of Walmart in our experiments.

Apart from sales, the M5 dataset also provides a list of covariates, including category information (the item, department, category, store, and
state ID of an SKU), calendar information (the date, week, day of week, month, event name and event type), prices of sales and SNAP
(Supplemental Nutrition Assistance Program) information. In this paper, we use these covariates in two of the individual methods described
in Table~\ref{tab:intro-individual}, i.e., LightGBM and LSTM, while the other seven individuals only use the historical sales data. In particular, GAM-QR uses averages of sales in the last 28 days and the last 7 days as input features. LightGBM and LSTM share the same features, using both historical sales and covariates. Sales features include sales in the last 28 days of the training set, and averages of sales in the last 28 days and last 7 days. Covariates include intermittence features (distances from the first day with positive sales, distances between the last two positive sales, and distances from the last positive sale), and ID features (store, state, category, department, and item ID), apart from those used in GAM-QR. For the ID features, we encode them as a number for LightGBM, while an embedding layer is used for LSTM to reduce dimensions as Table A.2 of the online Appendix A.

\subsection{Forecasting process} \label{process}

Aiming to obtain the 28-day-ahead probabilistic forecasts, we proceed with the following forecasting process.

Initially, the nine individual methods (see Table~\ref{tab:intro-individual}) are employed to yield individual forecasts for combinations. Note that they generate diverse outputs, ranging from quantiles to probabilistic distributions. To streamline probabilistic forecasting, we yield all quantiles for $\tau \in \{0.01,0.02,\cdots,0.99\}$ for each forecast horizon across all individual methods.

Second, to get the probability mass functions, we first round all the quantile forecasts and assign the $100\%$th quantiles equal to the $0.99$ quantiles. Then we calculate the frequencies of each value within the 100 quantiles as its probability. Eventually, for each individual method, quantile forecasts are transformed into the respective probabilistic distribution, which is utilized for combination. 

As a last step, combination weights are estimated based on the methods described in Section~\ref{comb}.

\subsection{Evaluation metrics}\label{metrics}

 Probabilistic forecasting is typically assessed concerning calibration and sharpness \citep{gneiting2014probabilistic,kolassa2016evaluating}, in contrast to point forecasting, which mainly focuses on accuracy. Additionally, intermittent demand forecasting is directly relevant to inventory management. Therefore, beyond calibration and sharpness, we extend to evaluate the inventory performance of forecasting, as advocated by \citet{goltsos2021inventory}. In a nutshell, our evaluation encompasses three key dimensions: calibration, sharpness, and inventory performance.

\subsubsection*{Calibration}

Calibration pertains to the statistical accuracy in the alignment between probabilistic forecasting and the corresponding realizations. In practice, the probability integral transform (PIT) histogram serves as a standard solution to diagnose the property of a distribution. For a continuous forecast distribution $\hat{F}_t$ and an observation $y_t$, PIT is
\begin{equation*}
    \text{PIT}_t = \hat{F}_t(y_t)=\int_{-\infty}^{y_t}\hat{f}_t(y)dy.
    \label{eq:pit}
\end{equation*} 
The objective is to test whether $\text{PIT}_t$ is on the uniform distribution $U[0,1]$. However, this form cannot be directly employed in discrete distributions, \citet{brockwell2007universal} introduces randomized PIT (rPIT) for discrete scenarios. \citet{kolassa2016evaluating} suggests drawing $pit_t$ from the distribution $U[\hat{F}_t(y_t-1),\hat{F}_t(y_t)]$. The $\text{PIT}_t$ will still be i.i.d on $U[0,1]$ if $\hat{F}_t$ is a proper distribution. In the paper, for each forecasting horizon, we sample 30 PIT values, yielding 840 in total for the 28-day-ahead forecasting. To evaluate the calibration of each combination method, we gather all PIT values from different series in the dataset and draw the corresponding histogram. 

\subsubsection*{Sharpness}
Sharpness measures the concentration of probabilistic forecasts and is only related to the forecasts themselves, with a sharper forecast considered superior. Sharpness can be evaluated by applying proper scoring rules \citep{gneiting2014probabilistic}. We employ the logarithmic score, DRPS score, and Brier score as introduced in Section \ref{comb} to evaluate sharpness. For comparison purposes, the logarithmic score is multiplied by $-1$, resulting in smaller scores representing sharper distributions. The logarithmic score is sensitive to close-to-zero probability for observations, underscoring the significance of forecast coverage for extreme situations. Brier score is linked to PMF directly, while DRPS score is associated with cumulative distribution function (CDF).

\subsubsection*{Inventory}
The summary evaluation of inventory management involves the calculation of simulated inventory costs during the forecast period, as detailed in Equation (\ref{eq:obj}). We use the ratio of the total simulated cost for all SKUs to the total sales in the 28-day horizon to gauge the economic outcomes of different methods. Then we analyze three crucial metrics, including the customer service level, lost sales, and inventory investment, drawing insights from \citet{syntetos2015forecasting} and \citet{trapero2019quantile}, to understand their trade-off relationship. The customer service level means the probability of not being out of stock at a specific target level (a quantile). Lost sales are the positive parts of real demand minus forecasting inventory at the given level during the period and averaged across SKUs, representing the average lack of sales per SKU. Inventory investment corresponds to the average forecasting inventory at target service levels across periods and SKUs, signifying the average maintained level of safety stocks. 

\subsection{Results}\label{res}

Here, we present the outcomes regarding calibration, sharpness, cost, and inventory performance, respectively. 

\begin{figure}[htbp]
  \centering
  \begin{minipage}[b]{0.32\textwidth}
  \centering
    \includegraphics[scale=0.2]{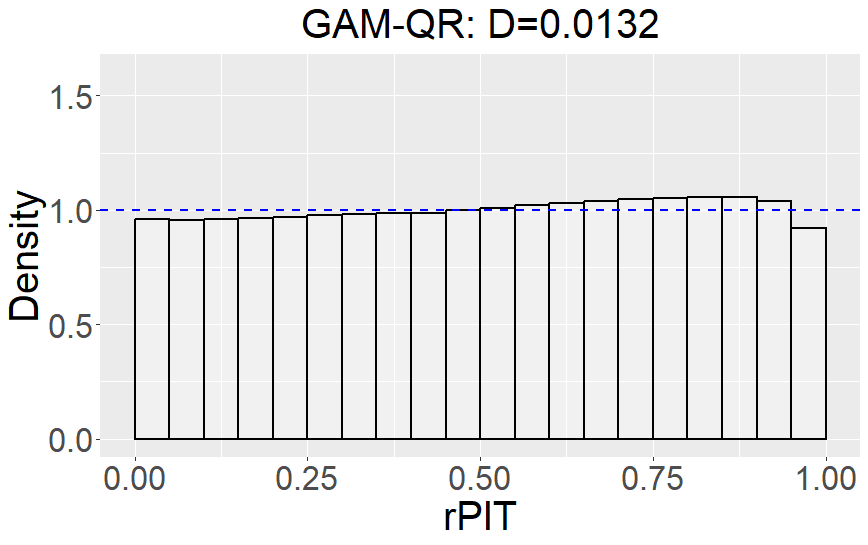}
  \end{minipage}
  \begin{minipage}[b]{0.32\textwidth}
  \centering
      \includegraphics[scale=0.2]{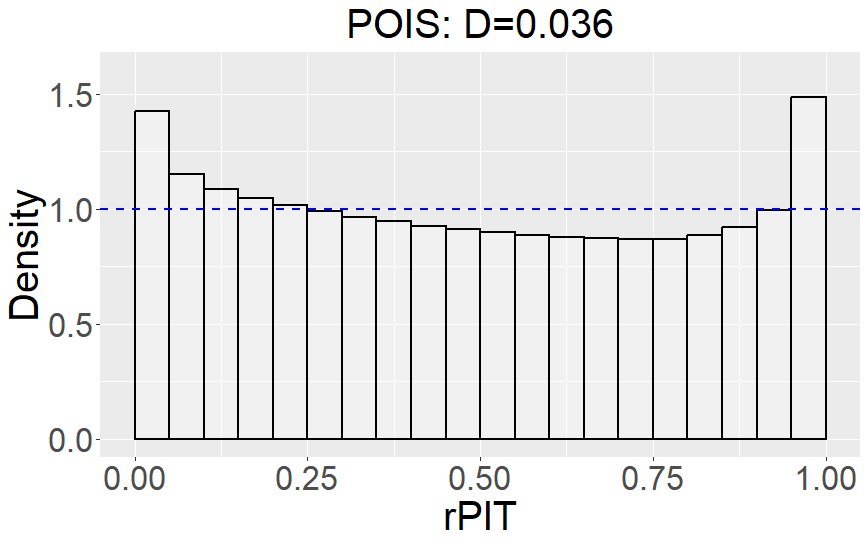}
  \end{minipage}
  \begin{minipage}[b]{0.32\textwidth}
  \centering
      \includegraphics[scale=0.2]{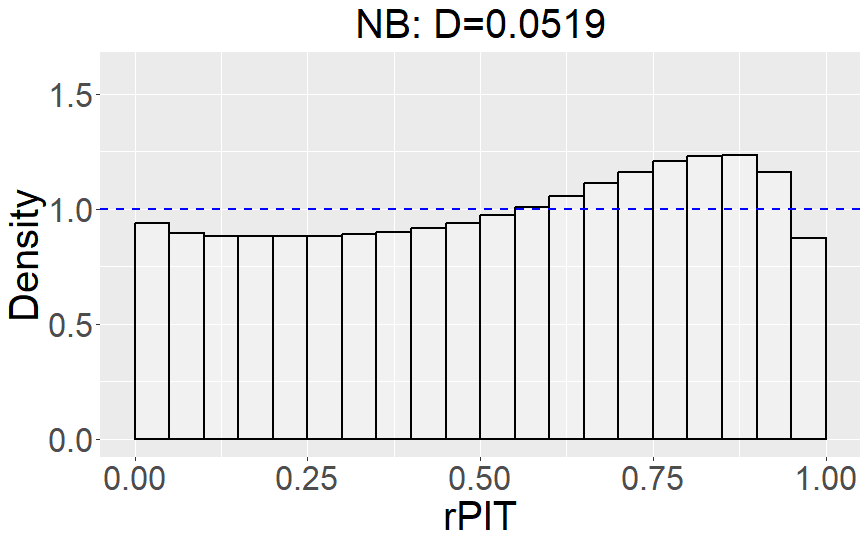}
  \end{minipage}
  \begin{minipage}[b]{0.32\textwidth}
  \centering
    \includegraphics[scale=0.2]{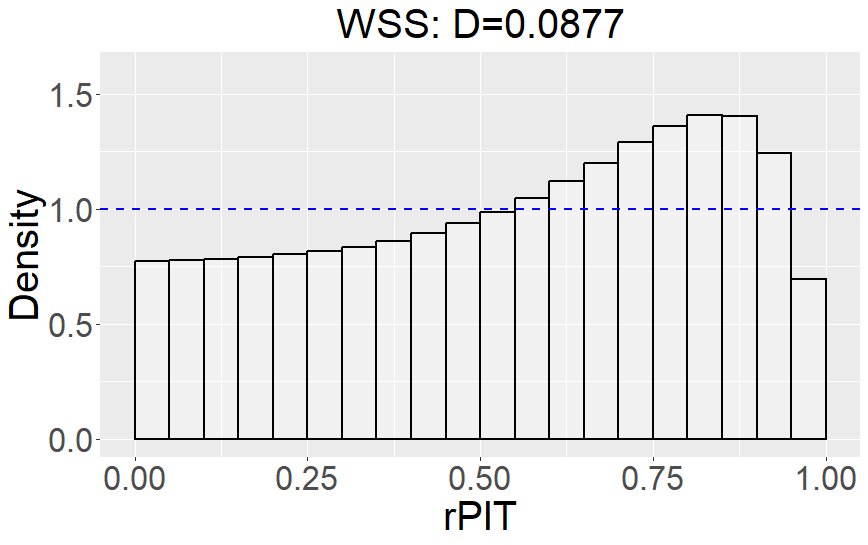}
  \end{minipage}
  \begin{minipage}[b]{0.32\textwidth}
  \centering
    \includegraphics[scale=0.2]{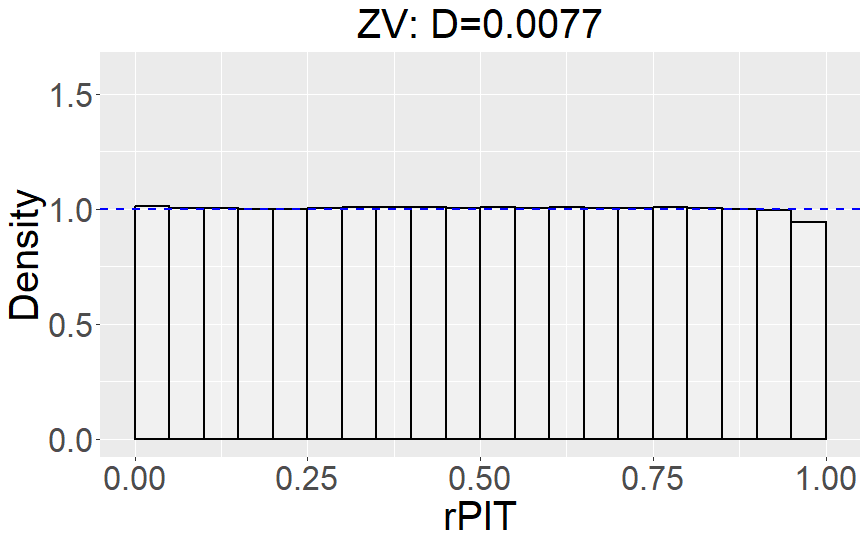}
  \end{minipage}
  \begin{minipage}[b]{0.32\textwidth}
  \centering
    \includegraphics[scale=0.2]{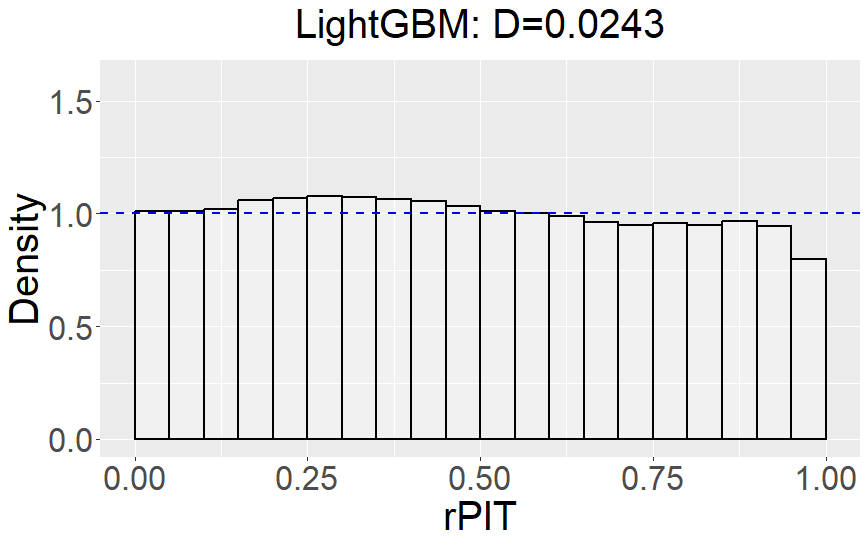}
  \end{minipage}
  \begin{minipage}[b]{0.32\textwidth}
  \centering
    \includegraphics[scale=0.2]{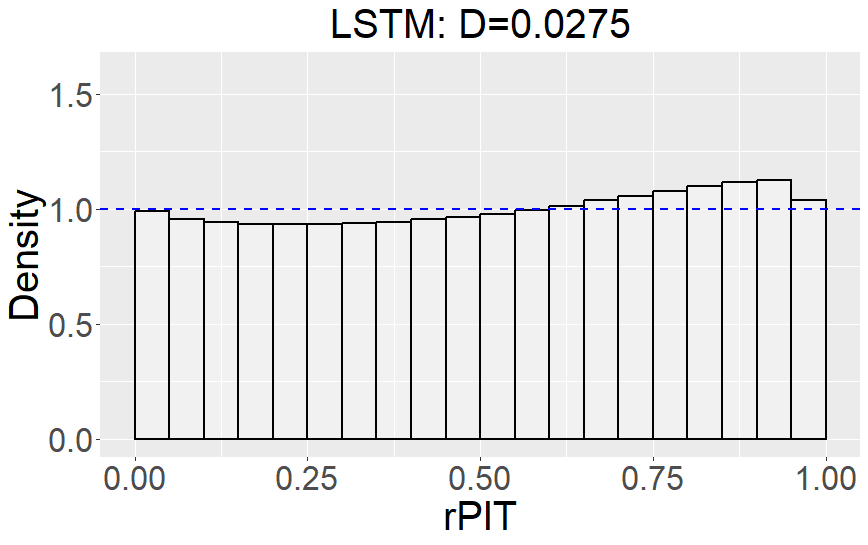}
  \end{minipage}
  \begin{minipage}[b]{0.32\textwidth}
  \centering
    \includegraphics[scale=0.2]{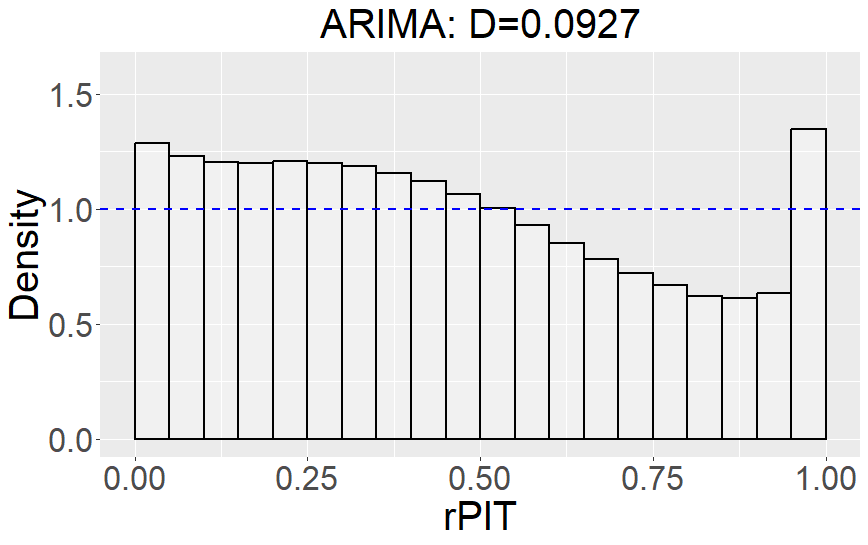}
  \end{minipage}
    \begin{minipage}[b]{0.32\textwidth}
  \centering
    \includegraphics[scale=0.2]{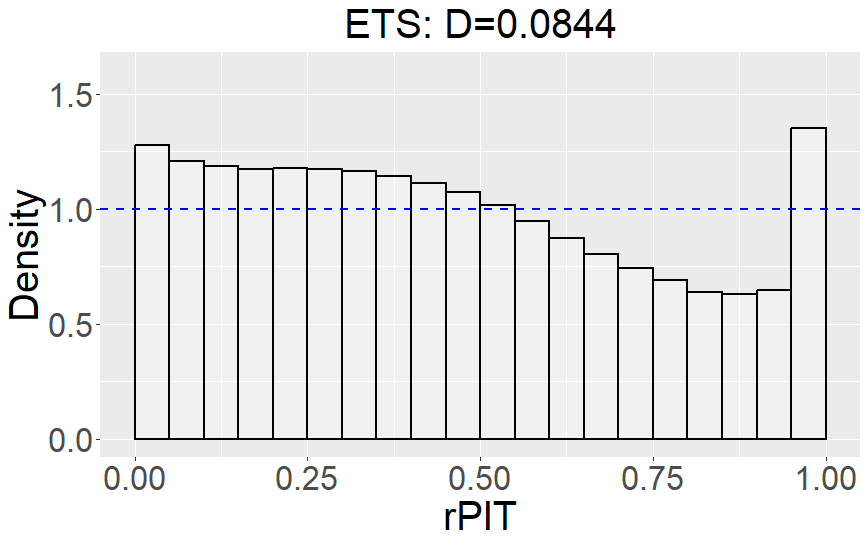}
  \end{minipage}
    
    \begin{minipage}[b]{0.32\textwidth}
  \centering
    \includegraphics[scale=0.2]{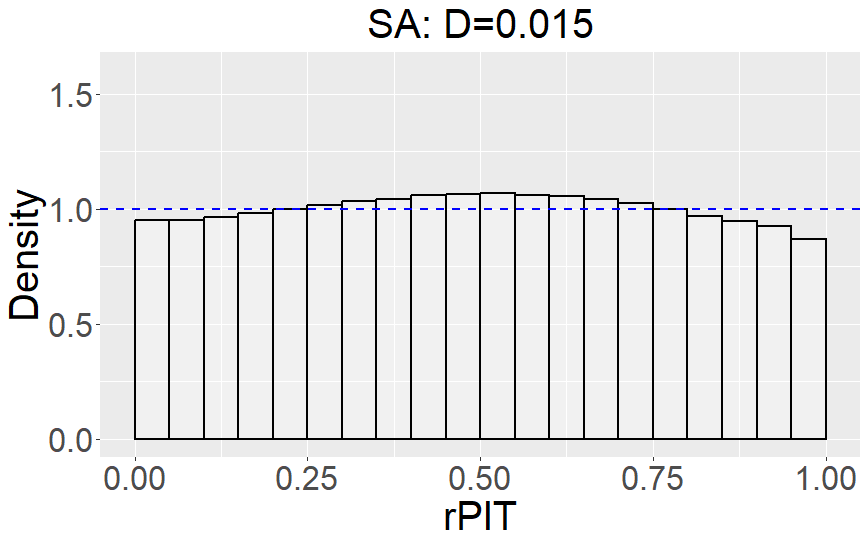}
  \end{minipage}
  \begin{minipage}[b]{0.32\textwidth}
  \centering
    \includegraphics[scale=0.2]{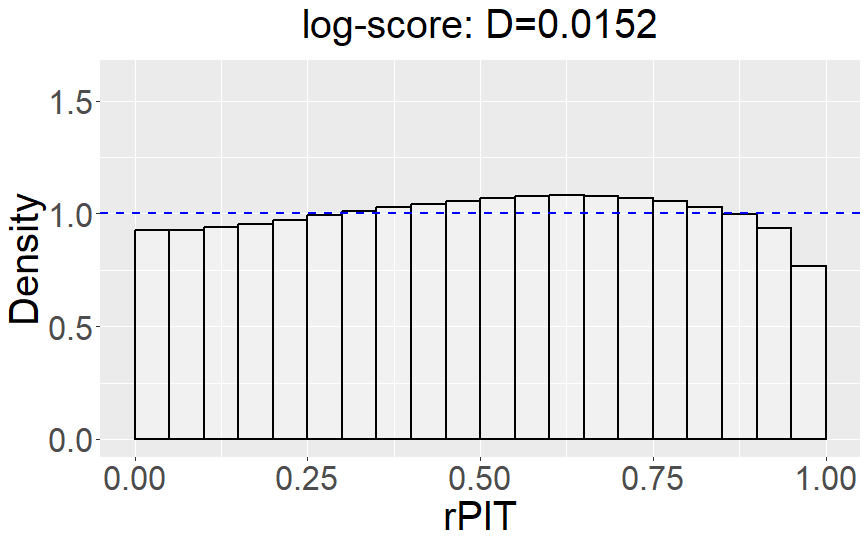}
  \end{minipage}
  \begin{minipage}[b]{0.32\textwidth}
  \centering
    \includegraphics[scale=0.2]{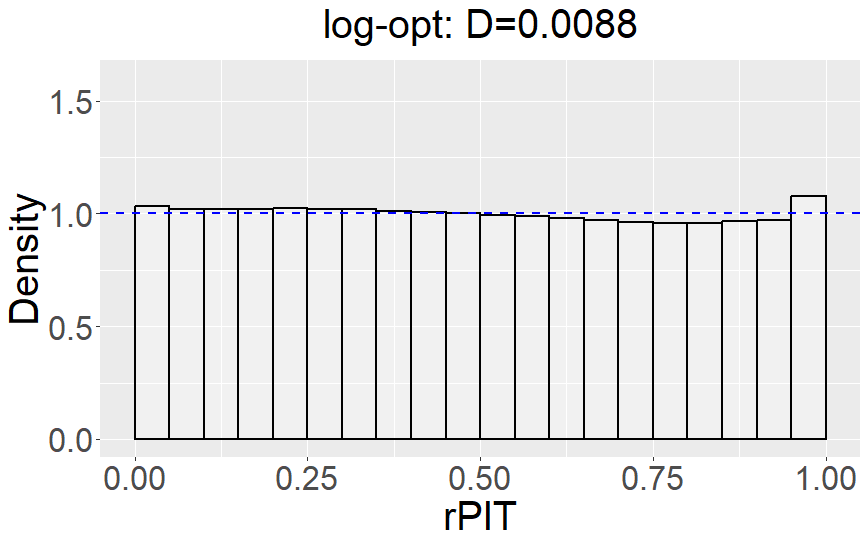}
  \end{minipage}
  \begin{minipage}[b]{0.32\textwidth}
  \centering
    \includegraphics[scale=0.2]{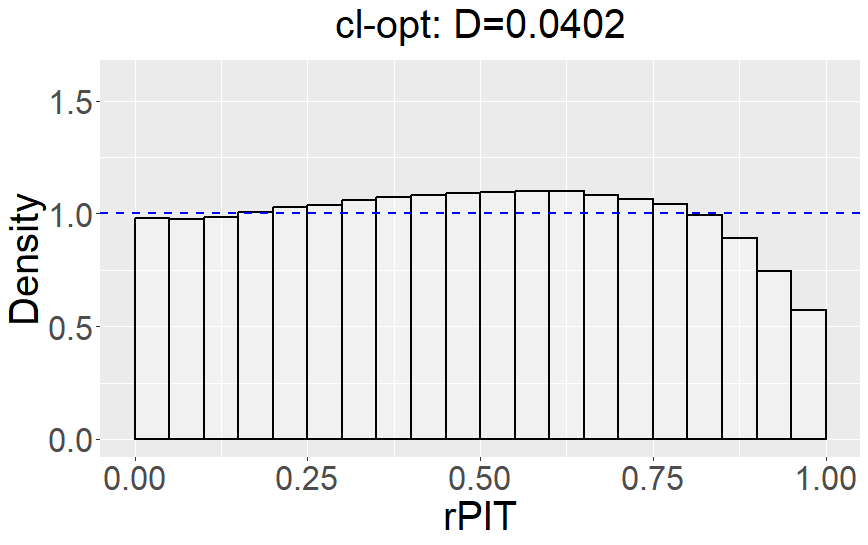}
  \end{minipage}
    \begin{minipage}[b]{0.32\textwidth}
  \centering
    \includegraphics[scale=0.2]{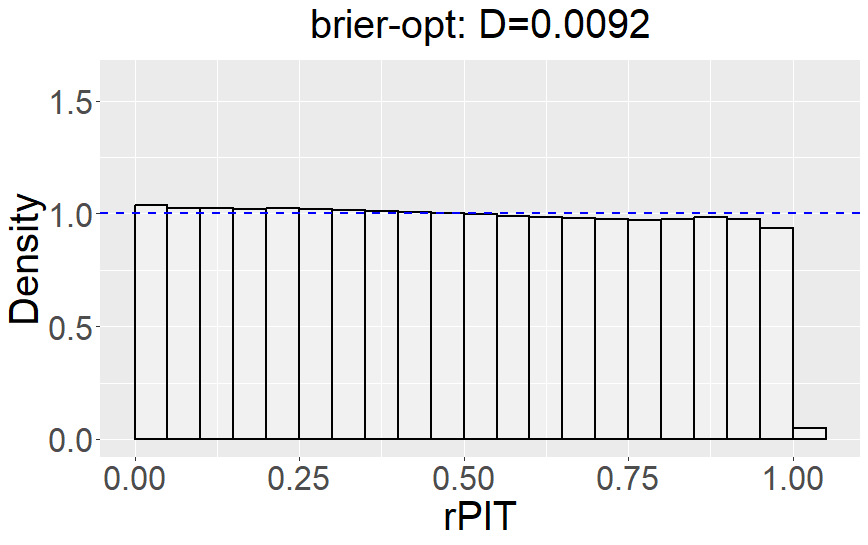}
  \end{minipage}
  \begin{minipage}[b]{0.32\textwidth}
  \centering
    \includegraphics[scale=0.2]{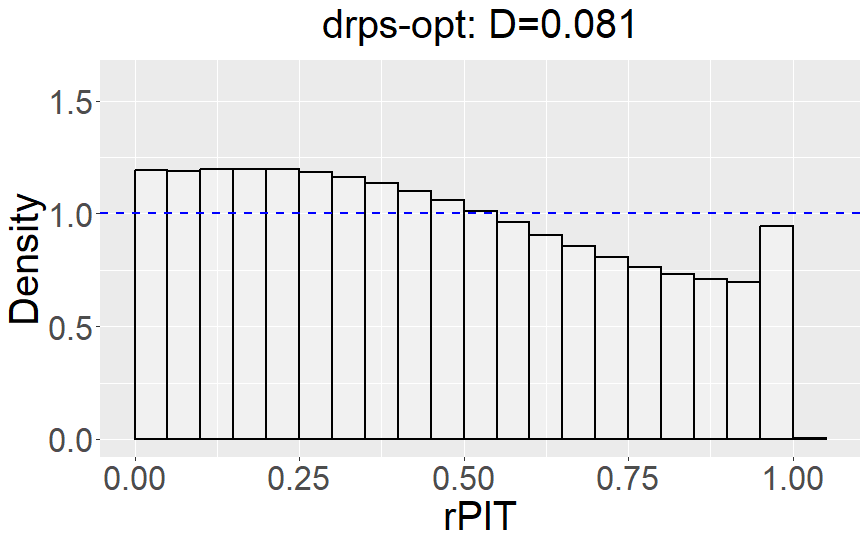}
  \end{minipage}
  \begin{minipage}[b]{0.32\textwidth}
  \centering
    \includegraphics[scale=0.2]{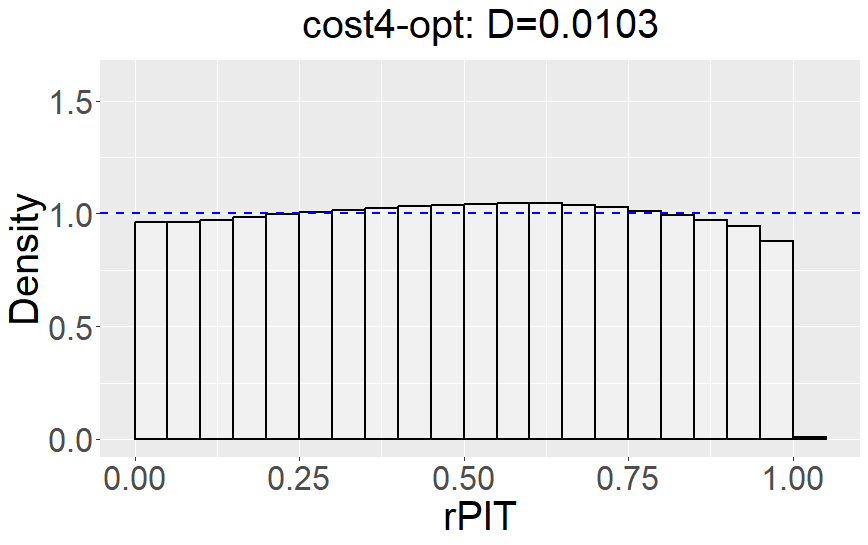}
  \end{minipage}
    \begin{minipage}[b]{0.32\textwidth}
  \centering
    \includegraphics[scale=0.2]{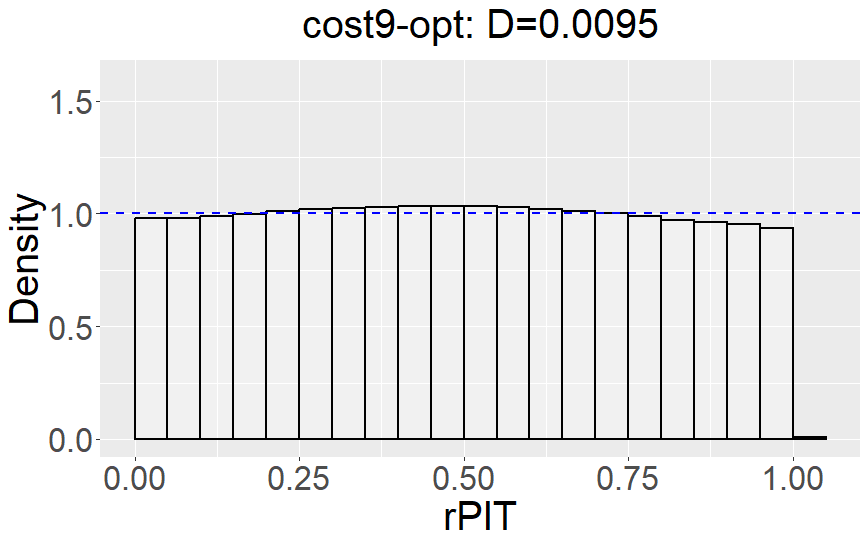}
  \end{minipage}
    \begin{minipage}[b]{0.32\textwidth}
  \centering
    \includegraphics[scale=0.2]{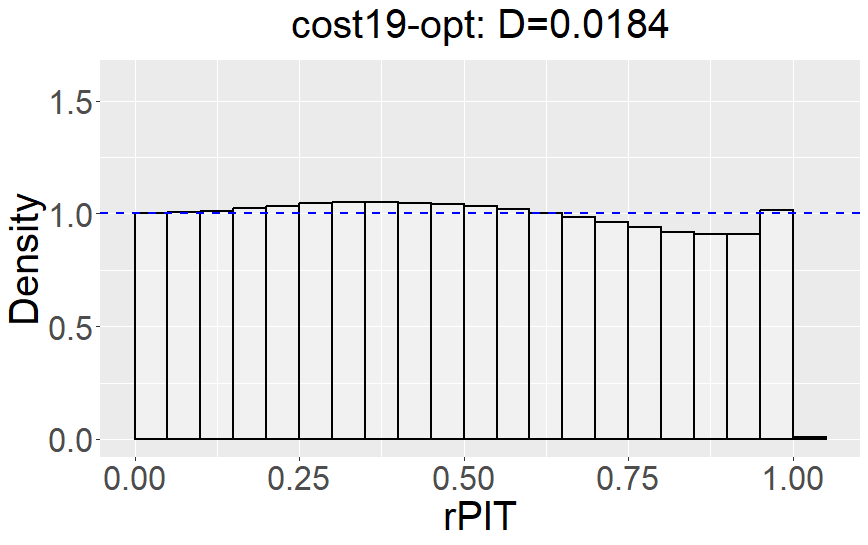}
  \end{minipage}

  \caption{The rPIT histograms of all individual and combination methods. The dot lines mark the ideal uniform distribution, and the bar heights represent the real densities. $D$ refers to the statistics of Kolmogorov-Smirnov test, quantifying the distance between sample distribution to the uniform distribution $U[0,1]$.}
  \label{fig:calibration of dataset}
\end{figure}

The calibration performance of all methods is shown in Figure \ref{fig:calibration of dataset}. We show in each subplot the Kolmogorov-Smirnov (KS) test statistic $D$ \citep{massey1951kolmogorov}, indicating the maximum difference between sample distribution and theoretical distribution ($U[0,1]$ in our case)~. Across the nine individual methods, the results vary considerably. ZV, GAM-QR, and LightGBM are more calibrated than others. GAM-QR and POIS display peaks at both tails, indicating an underestimation of the distribution in extreme scenarios such as zero-sale and high sales. WSS and NB tend to underestimate the distribution of higher sales but overestimate extremely high sales. Conversely, ARIMA and ETS overestimate the high part and underestimate the low part of the distribution. Turning to combination methods, both simple weighting methods (SA and log-score), log-opt, brier-opt, and cost-based methods with low back-order marginal costs (cost4-opt and cost9-opt) demonstrate effective calibration. Overall, combining individual probabilistic forecasts yields more calibrated results, overcoming the problem of underestimating high sales in some individual forecasts. 

The sharpness results are presented in Table \ref{tab:sharpness}. Among combination methods, SA outperforms others in the logarithmic and Brier score, while drps-opt is outstanding in the DRPS score. Notably, combinations demonstrate a significant improvement over individual methods across all three scoring rules. However, for scoring-based combinations, achieving an optimal result proves challenging, even when the optimal objectives align with the evaluation metric. For example, drps-opt realizes the best performance on the DRPS score, while brier-opt and log-opt are inferior to SA in the Brier and logarithmic scores, respectively. Compared with scoring-based combinations, SA emerges as a fair and robust choice across all metrics. Furthermore, We observe that the rankings of the logarithmic and Brier scores correlate slightly positively while the DRPS score is inversely correlated with the other two. These results are reasonable. Firstly, as discussed by \citet{buja2005loss} and \citet{merkle2013choosing}, the logarithmic score and Brier score share a family of scoring rules. Besides, DRPS focuses on CDF while the other two evaluate forecasts based on PMF.

\begin{table}[!ht]
\centering
\caption{The sharpness performance regarding three proper scoring rules, i.e., logarithmic score, DRPS score and Brier score. The bold numbers represent the best method in the metrics of its group (individuals and combinations).}
\label{tab:sharpness}
{\begin{tabular}{lcccclccc}
\toprule
\multicolumn{4}{c}{Individuals} & & \multicolumn{4}{c}{Combinations}\\
    \cline{1-4} \cline{6-9} 
Method & Logarithmic & DRPS & Brier & & Method & Logarithmic & DRPS & Brier\\ \midrule 
GAM-QR & 4.375& 1.213 & -0.411 & &SA & \textbf{1.706} & 1.124 & \textbf{-0.420}\\
POIS & 7.434 & 1.140 & -0.409& &log-score & 1.713 & 1.140 & -0.418 \\
NB & 3.393 & 1.246 & -0.404 & &log-opt & 1.787 & 1.129 & -0.416 \\
WSS & \textbf{2.961} & 1.347 & -0.374& &cl-opt & 1.790& 1.135 & -0.405 \\
ZV & 3.714 & 1.187 & -0.410 && brier-opt & 1.754 & 1.128 & -0.415 \\
LightGBM & 3.954 & 1.156 & -0.400 && drps-opt & 1.800 & \textbf{1.014} & -0.400\\
LSTM & 3.902 & 1.208 & \textbf{-0.412}&& cost4-opt & 1.891 & 1.133 & -0.418 \\
ARIMA & 8.444 & \textbf{1.027} & -0.391 && cost9-opt & 1.998 & 1.130 & -0.418\\
ETS & 8.395 & 1.034 & -0.394 && cost19-opt & 2.209 & 1.115 & -0.416\\
\bottomrule 
\end{tabular}}
\end{table}

Table \ref{tab:cost} shows the costs under various lost-sale cost conditions, including cost(1,4), cost(1,9), and cost(1,19). The results of cost-opt illustrate the costs of cost4-opt for cost(1,4), cost9-opt for cost(1,9), and cost19-opt for cost(1,19). 
For individuals, ETS is superior to others under the three cost conditions. For combinations, cost-opt surpasses both individuals and combinations across different conditions. SA is only inferior to cost-opt across the three cost conditions. Overall, combinations demonstrate the capability to achieve lower costs compared to individual methods, with both cost-opt and SA emerging as highly promising approaches for reducing inventory costs.

\begin{table}[!ht]
\centering
\caption{The simulated cost of all individual and combination methods.}
\label{tab:cost}
{\begin{tabular}{llcccc}
\toprule
Type & Method & cost(1,4) & cost(1,9) & cost(1,19)   \\
\midrule 
\multirow{9}*{Individuals} & GAM-QR & 1.693 & 2.396 & 3.104  \\
~ & POIS  & \textbf{1.551} & 2.260 & 3.112  \\
~ & NB  & 1.613 & 2.338 & 3.162  \\
~ & WSS  & 1.765 & 2.505 & 3.315  \\
~ & ZV  & 1.679 & 2.400 & 3.160  \\
~ & LightGBM  & 1.711 & 2.466 & 3.262  \\
~ & LSTM  & 1.612 & 2.286 & 3.009  \\
~ & ARIMA & 1.589 & 2.222 & 2.956  \\
~ & ETS  & 1.556 & \textbf{2.174} & \textbf{2.891}  \\
\hline
\multirow{7}*{Combinations} & SA & 1.522 & 2.168 & 2.863  \\
~ & log-score  & 1.536 & 2.205 & 2.912  \\
~ & log-opt  & 1.534 & 2.178 & 2.860 \\
~ & cl-opt & 1.675 & 2.464 & 3.266 \\
~ & brier-opt & 1.540 & 2.193 & 2.896  \\
~ & drps-opt  & 1.550 & 2.194 & 2.909  \\
~ & cost-opt  & \textbf{1.517} & \textbf{2.146} & \textbf{2.826} \\
\bottomrule 
\end{tabular}}
\end{table}

\begin{figure}[htbp]
\centering
\includegraphics[width=1\linewidth]{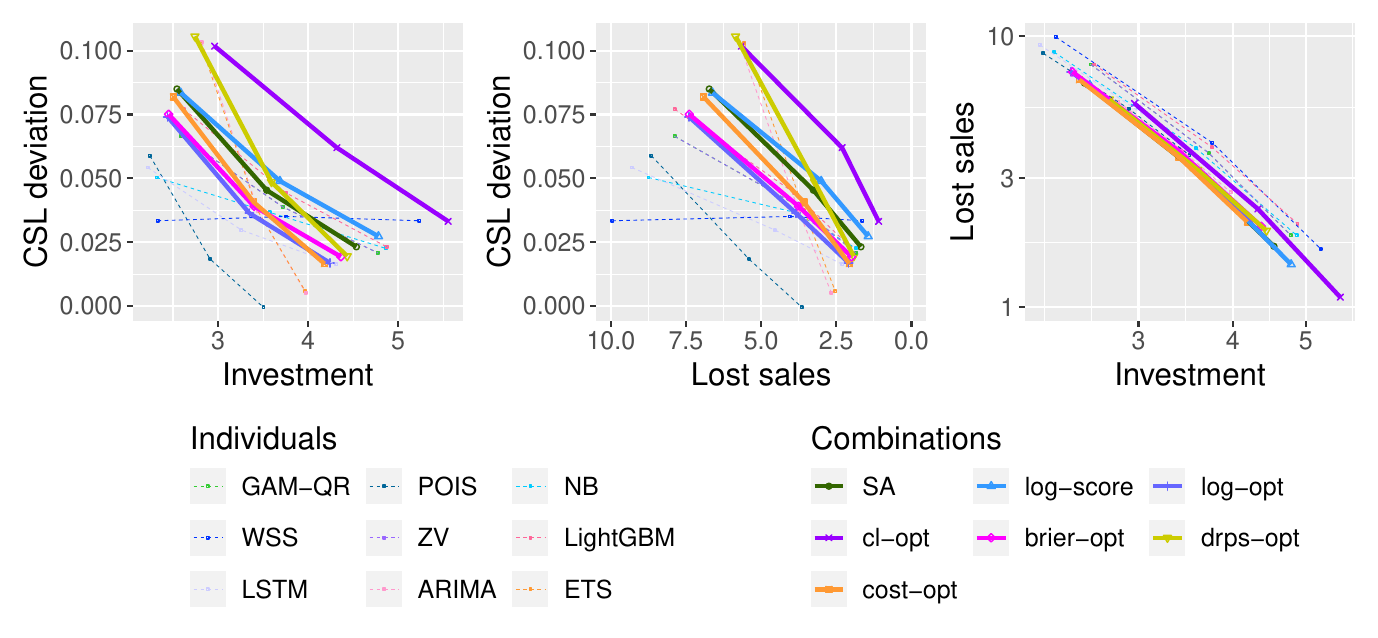}
  \caption{The relationship among CSL, inventory investment, and lost sales. The three points on the lines represent the target levels of 80\%, 90\%, and 95\%. CSL deviation is the gap in customer service levels between the achieved one and target one. Investment is the average of stocks each day over all SKUs and all horizons. ``Lost sales'' in the figures mean the sum of lost sales in all the horizons averaged by SKUs. Logarithmic axes are in the right subplot for visualization purposes, making the curve shapes concave rather than convex in common axes.}
  \label{fig:inventory}
\end{figure}

Aiming at delve deeper into inventory effects, Figure \ref{fig:inventory} illustrates the relationship among the deviations between target and achieved customer service levels (the achieved ones minus target ones, named CSL deviation), the average maintained daily investment and the cumulative lost sales over the 28 forecasting horizons. Ideally, our aspiration for the proposed methods is zero CSL deviation, low investment, and zero lost sales.  Seven combinations are distinctly represented with thick and solid lines to accentuate the variances between combinations and individuals. The employed inventory policy is described in Section \ref{comb-cost}. Our focal target service levels are 0.8, 0.9 and 0.95, corresponding to the three points lying from left to right of the lines. We match the results of cost4-opt, cost9-opt, and cost19-opt to the three levels, respectively. 

In the left panel of Figure~\ref{fig:inventory}, log-opt emerges as the combination with the lowest investment across all target levels. Furthermore, it exhibits smaller deviations from target levels compared to other combinations. Among all individual forecasts, POIS stands out as superior in both CSL deviation and investment. However, the trade-off relationship between investment and lost sales is evident across all methods. In the middle panel, POIS records the largest lost sales, with log-opt and brier-opt also leading to more lost sales than other combination methods. The right panel further tells the trade-off relationship. SA, log-score and cost-opt are nearer to the origin than the rest, i.e., they have a fair balance between the two metrics. However, cost-opt cannot adapt to high target service levels like 95\%, meaning that it reduces investment to decrease costs. Overall, SA performs more robustly in balancing investment and lost sales.

Our exploration extends to the performances across various intermittent patterns. Following the categorization proposed by \citet{syntetos2005categorization}, intermittent demand is classified into four patterns based on the average demand interval (ADI) and the square of the coefficient of variation (CV2), denoted as ``smooth'', ``erratic'', ``lumpy'', and ``intermittent''. The critical values of segmentation are 1.32 for ADI and 0.49 for CV2. Among the 28903 time series, the series numbers of the four patterns are 1994, 811, 4987, and 21111. Details about sharpness and cost evaluation results are shown from Table C.4 to Table C.11 in the online Appendix C. Our findings can be summarized as follows. First, the trade-off relationship between forecasting and inventory exists in all four demand patterns. Second,  the rankings of methods exhibit minimal variation across different demand patterns, whether in terms of sharpness or inventory. Third, the intermittent category has the lowest sharpness scores among the four demand patterns, while we see higher costs in the lumpy and intermittent groups. Collectively, these insights contribute to a nuanced understanding of how different forecasting methods perform under varying intermittent demand scenarios.

Furthermore, an examination of the influences of covariates on GAM-QR, LightGBM and LSTM is presented in Table B.3 of the online Appendix B. Regarding GAM-QR, the ``noco'' version outperforms the ``co'' version across both sharpness and cost. The phenomenon is opposite to LSTM. For LightGBM, although the ``noco'' version demonstrates slightly better in sharpness, it incurs more cost in inventory. That is also why we use the non-covariate version of GAM-QR in combinations, while including covariates for LightGBM and LSTM.

In a nutshell, the amalgamation of probabilistic forecasts presents notable enhancements in both forecasting and inventory management. However, the use of approaches like ETS and ARIMA that choose the best out of a set of possible models still performs reasonably well and, in some cases, better than probabilistic forecast combinations. In any case, a trade-off relationship exists between the two tasks. For example, while log-opt excels in forecasting assessment using the logarithmic scoring rule, its superior forecasting performance does not necessarily translate into cost savings in inventory management. Finally, SA emerges as a method capable of achieving a balanced and equitable outcome across both forecasting and inventory management tasks.

\section{Conclusion}\label{conclusion}

This paper proposed a novel approach to probabilistic forecasting combinations for intermittent demand, emphasizing a holistic consideration of both forecasting accuracy and inventory performance. Our goal is to synthesize forecasting and inventory management for more effective decision-making. Our evaluation of the forecasting accuracy and economic performance of these combinations reveals that they outperform individual methods. However, a trade-off between forecasting and inventory performance is evident, wherein the best-performing method in terms of calibration and sharpness may not necessarily result in reduced inventory costs, and vice versa. Therefore, forecasting performance should not be the sole factor in inventory evaluation. Remarkably, the simple average method achieves outstanding results in both tasks and exhibits robustness, while other combinations based on optimization perform well in only one target.

This research provides three important insights into intermittent demand forecasting. Firstly, probabilistic forecasting needs to be addressed in the field, and probabilistic combinations are helpful as they have demonstrated success in other areas. Secondly, we should consider forecasting and inventory together to fully capture the practical values of intermittent demand, including both evaluation and strategy construction. Finally, the simple average method emerges as advantageous for probabilistic forecasting of intermittent demand, particularly when evaluating inventory performance. This finding reminds us of the ``forecast combination puzzle'' \citep{stock2004combination}. \citet{CLAESKENS2016754} gave a theoretical explanation on the issue, implying that the simple average may be a solid benchmark. 

\section*{Acknowledgments}

We are grateful to Editor and the two anonymous reviewers for their insightful comments and suggestions, which significantly improved the paper. Yanfei Kang is supported by the National Natural Science Foundation of China (No. 72171011). This research was supported by the high-performance computing (HPC) resources at Beihang University.

\bibliographystyle{model5-names}
\bibliography{main}

\end{document}